\documentclass[sigconf]{acmart}
\usepackage{amsmath,amsfonts}
\usepackage{algorithm}
\usepackage{array}
\usepackage[caption=false,font=normalsize,labelfont=sf,textfont=sf]{subfig}
\usepackage{textcomp}
\usepackage{stfloats}
\usepackage{verbatim}
\usepackage{graphicx}
\usepackage{adjustbox}
\usepackage{booktabs}
\usepackage{tabularx}
\usepackage{multicol}
\usepackage[bottom]{footmisc}

\usepackage{algpseudocode}
\usepackage{amsthm,amssymb,lipsum}
\usepackage{cleveref}
\usepackage{caption}
\usepackage{url}
\usepackage{ulem}
\usepackage{threeparttable}
\crefname{section}{§}{§§}
\usepackage{listings, xcolor}
\definecolor{verylightgray}{rgb}{.97,.97,.97}
\usepackage{fancyhdr}

\lstdefinelanguage{Solidity}{
	keywords=[1]{anonymous, assembly, assert, balance, break, call, callcode, case, catch, class, constant, continue, constructor, contract, debugger, default, delegatecall, delete, do, else, emit, event, experimental, export, external, false, finally, for, function, gas, if, implements, import, in, indexed, instanceof, interface, internal, is, length, library, log0, log1, log2, log3, log4, memory, modifier, new, payable, pragma, private, protected, public, pure, push, require, return, returns, revert, selfdestruct, send, solidity, storage, struct, suicide, super, switch, then, this, throw, transfer, true, try, typeof, using, value, view, while, with, addmod, ecrecover, keccak256, mulmod, ripemd160, sha256, sha3}, 
	keywordstyle=[1]\color{blue}\bfseries,
	keywords=[2]{address, bool, byte, bytes, bytes1, bytes2, bytes3, bytes4, bytes5, bytes6, bytes7, bytes8, bytes9, bytes10, bytes11, bytes12, bytes13, bytes14, bytes15, bytes16, bytes17, bytes18, bytes19, bytes20, bytes21, bytes22, bytes23, bytes24, bytes25, bytes26, bytes27, bytes28, bytes29, bytes30, bytes31, bytes32, enum, int, int8, int16, int24, int32, int40, int48, int56, int64, int72, int80, int88, int96, int104, int112, int120, int128, int136, int144, int152, int160, int168, int176, int184, int192, int200, int208, int216, int224, int232, int240, int248, int256, mapping, string, uint, uint8, uint16, uint24, uint32, uint40, uint48, uint56, uint64, uint72, uint80, uint88, uint96, uint104, uint112, uint120, uint128, uint136, uint144, uint152, uint160, uint168, uint176, uint184, uint192, uint200, uint208, uint216, uint224, uint232, uint240, uint248, uint256, var, void, ether, finney, szabo, wei, days, hours, minutes, seconds, weeks, years},	
	keywordstyle=[2]\color{teal}\bfseries,
	keywords=[3]{block, blockhash, coinbase, difficulty, gaslimit, number, timestamp, msg, data, gas, sender, sig, value, now, tx, gasprice, origin},	
	keywordstyle=[3]\color{violet}\bfseries,
	identifierstyle=\color{black},
	sensitive=true,
	comment=[l]{//},
	morecomment=[s]{/*}{*/},
	commentstyle=\color{gray}\ttfamily,
	stringstyle=\color{red}\ttfamily,
	morestring=[b]',
	morestring=[b]"
}

\usepackage{tcolorbox}

\AtBeginDocument{%
  \providecommand\BibTeX{{%
    \normalfont B\kern-0.5em{\scshape i\kern-0.25em b}\kern-0.8em\TeX}}}

\copyrightyear{2024}
\acmYear{2024}
\setcopyright{acmlicensed}
\acmConference[WWW '24 Companion] {Companion Proceedings of the ACM Web Conference 2024}{May 13--17, 2024}{Singapore, Singapore}
\acmBooktitle{Companion Proceedings of the ACM Web Conference 2024 (WWW '24 Companion), May 13--17, 2024, Singapore, Singapore}
\acmDOI{10.1145/3589335.3651562}
\acmISBN{979-8-4007-0172-6/24/05}
\begin{document}

\title{\textbf{StateGuard: Detecting State Derailment Defects in Decentralized Exchange Smart Contract}}

\author{Zongwei Li}

\affiliation{%
  \institution{Hainan University}
  \city{Haikou}
  \country{China}
}
\email{lizw1017@gmail.com}

\author{Wenkai Li}
\affiliation{%
  \institution{Hainan University}
  \city{Haikou}
  \country{China}
}
\email{liwenkai871@gmail.com}

\author{Xiaoqi Li}
\authornote{Corresponding author}
\affiliation{%
  \institution{Hainan University}
  \city{Haikou}
  \country{China}
}
\email{csxqli@gmail.com}

\author{Yuqing Zhang}
\affiliation{%
  \institution{University of Chinese Academy of Sciences}
  \city{Beijing}
  \country{China}
}
\email{zhangyq@nipc.org.cn}

\renewcommand{\shortauthors}{Zongwei Li, Wenkai Li, Xiaoqi Li \& Yuqing Zhang}

\begin{abstract}
Decentralized Exchanges (DEXs), leveraging blockchain technology and smart contracts, have emerged in decentralized finance. However, the DEX project with multi-contract interaction is accompanied by complex state logic, which makes it challenging to solve state defects. In this paper, we conduct the first systematic study on state derailment defects of DEXs. These defects could lead to incorrect, incomplete, or unauthorized changes to the system state during contract execution, potentially causing security threats. We propose \textsc{StateGuard}, a deep learning-based framework to detect state derailment defects in DEX smart contracts. \textsc{StateGuard} constructs an Abstract Syntax Tree (AST) of the smart contract, extracting key features to generate a graph representation. Then, it leverages a Graph Convolutional Network (GCN) to discover defects. Evaluating \textsc{StateGuard} on 46 DEX projects with 5,671 smart contracts reveals its effectiveness, with a precision of 92.24\%. To further verify its practicality, we used \textsc{StateGuard} to audit real-world smart contracts and successfully authenticated multiple novel CVEs.
\end{abstract}




\maketitle
\thispagestyle{empty} 
\fancyfoot{} 

\vspace{-0.4em}
\section{Introduction}
\vspace{-1ex}

DEXs are at the forefront of a trading revolution, enabling peer-to-peer transactions without the need for traditional intermediaries~\cite{Ramseyer_2023_SPEEDEXa}. In contrast to Centralized Exchanges (CEX), DEX avoids traditional custodial responsibilities and allows users to maintain complete control of their assets throughout the trading process. This reduces user risk from exchange misconduct or bankruptcy. DEXs have become a core component of decentralized finance (DeFi) but have encountered some challenges~\cite{LI2022DefiSurvey,li2020characterizing,li2021clue,zhang2022authros}. Despite security advantages, DEX platforms remain vulnerable to various threats, such as fund theft, market manipulation, and denial of service attacks\cite{Xia_2021_Tradea}.

However, despite several studies~\cite{Ramseyer_2023_SPEEDEXa, Xia_2021_Tradea, Duan_2022_Automated, Li_2021_SolSaviour} dedicated to revealing defects in DEX projects, detecting defects in the smart contracts of these projects remains challenging, mainly due to the following factors:

\textbf{Challenge 1 (C1)}: Related to the complex state logic of DEX projects on the Ethereum blockchain, the numerous transactions and interactions may change state unpredictably. Attackers can exploit state changes to trigger errors that are difficult to detect.

\textbf{Challenge 2 (C2)}: State derailment defects, which are different from common defects~\cite{otherdefects}. They stem from logic errors, resource constraints, access control errors, type and declaration errors, and mishandling of exceptions. These defects can cause unauthorized or incorrect state modifications, leading to anomalous behaviors or security threats during smart contract execution.

To address these challenges, we propose a deep learning-based framework called \textsc{StateGuard}, specifically designed to analyze and detect state derailment defects in DEX project smart contracts. For C1, \textsc{StateGuard} converts smart contracts into ASTs to capture key logical structural features, including five dependency features critical for state derailment: declaration dependencies, expression dependencies, control dependencies, data dependencies, and function dependencies. These ASTs map the state patterns, structures, and syntax of smart contracts in detail at the semantic level. For C2, the normalized data is further processed using a GCN model. This model identifies state derailment defects by learning defective features. GCN can efficiently process the attribute information of nodes and the connectivity relationship, combining node attributes and topology information to capture malicious behaviors and patterns in DEX smart contracts.

The main contributions of this paper are as follows:

\begin{itemize}
\vspace{-0.3em}
\item To the best of our knowledge, we present the first systematic study of state defects in DEX smart contracts. We define a new kind of state derailment defect, which can lead to unauthorized or incorrect modifications to the system state during the execution of smart contracts (\cref{sec:State Derailment Defects}).

\item We propose \textsc{StateGuard}, a novel deep learning-based framework for detecting and analyzing state defects in DEX projects. It learns structural features from the ASTs of DEX contracts and extracts five dependent features to identify state derailment defects (\cref{sec:methodology}). 

\item We comprehensively evaluated \textsc{StateGuard} on 46 DEX projects and 5,671 smart contracts with 94.25\% F1-score. We conduct a comparative analysis with state-of-the-art, with the advantages of 7.39\% in F1-score and 22.31\% in accuracy. In addition, \textsc{StateGuard} has discovered multiple novel real-world defects CVE-2023-\{47033, 47034, 47035\} (\cref{sec:experiment}).

\item We open-source \textsc{StateGuard}'s codes and experimental data on \url{https://doi.org/10.6084/m9.figshare.24715650}.

\end{itemize}

\vspace{-1em}
\section{Related Work} \label{sec:related work}
\vspace{-1ex}

Users exchange assets directly in DEX through smart contracts, and the security of smart contracts is crucial to user assets and needs to be comprehensively analyzed. Such a comprehensive analysis typically involves auditing DEX smart contracts, evaluating the robustness of their design, and modeling potential attack scenarios ~\cite{Zheng_2023_Blockchain-Based}. Since DEX operates on the blockchain, any security defect can be exploited by malicious attackers, which may lead to significant financial losses. Duan et al.~\cite{Duan_2022_Automated} proposed VetSC, an automated tool designed to perform security analysis of DApps. Meanwhile, Li et al.~\cite{Li_2021_SolSaviour} developed SolSaviour, a system that utilizes a voting-based mechanism to fix defective smart contracts. Moreover, Xia et al.~\cite{Xia_2021_Tradea} employed a machine-learning approach to detect fraud on Uniswap, And Geoffrey et al. ~\cite{Ramseyer_2023_SPEEDEXa} introduced SPEEDEX, a system designed to combat front-loading and increase the processing power of exchanges. Overall, the security analysis of DEX is a complex, multi-dimensional, and multi-layered process that requires a deep understanding of technical details, market behavior, and potential attack patterns.

\vspace{-0.8em}
\section{State Derailment Defects} \label{sec:State Derailment Defects}
\vspace{-1ex}
State derailment defects in DEX smart contracts are security defects that arise from improperly modifying the system state of a contract. The system state is a critical component that encapsulates the stored information or variables representing the current status or condition of the smart contract. 
A state derailment defect indicates an unauthorized, incorrect, or incomplete alteration to the system state due to logical inconsistencies, design flaws, resource constraints, or other unforeseen issues within the smart contract's code. Such defects might lead to abnormal system operations and allow malicious actors to exploit these defects, compromising the integrity and security of the smart contract state.

Fig. \ref{fig:code2} illustrates a state derailment defect in the ERC20 token standard. In this code, the \texttt{safeTransferFrom} function assumes the crucial role of transferring tokens between two addresses. Despite this, the function is set to \texttt{public} and lacks sufficient validation measures to ensure the state's legitimacy. Any user can call this function, potentially triggering a state derailment defect.

\begin{figure}[!t]
\setlength{\abovecaptionskip}{0.cm}
\begin{lstlisting}[numbers=none]
function safeTransferFrom(IERC20Token _token, address _from, address _to, uint256 _value) public {
   execute(_token, abi.encodeWithSelector(TRANSFER_FROM_FUNC_SELECTOR, _from, _to, _value));
}
\end{lstlisting}
\caption{Code Snippets of Defective Contracts.}
\label{fig:code2}
\vspace{-1.2em}
\end{figure}

\section{Methodology} \label{sec:methodology}
\vspace{-1ex}

\begin{figure}[!t]
\vspace{-2em}
\centering
\includegraphics[width=\linewidth]{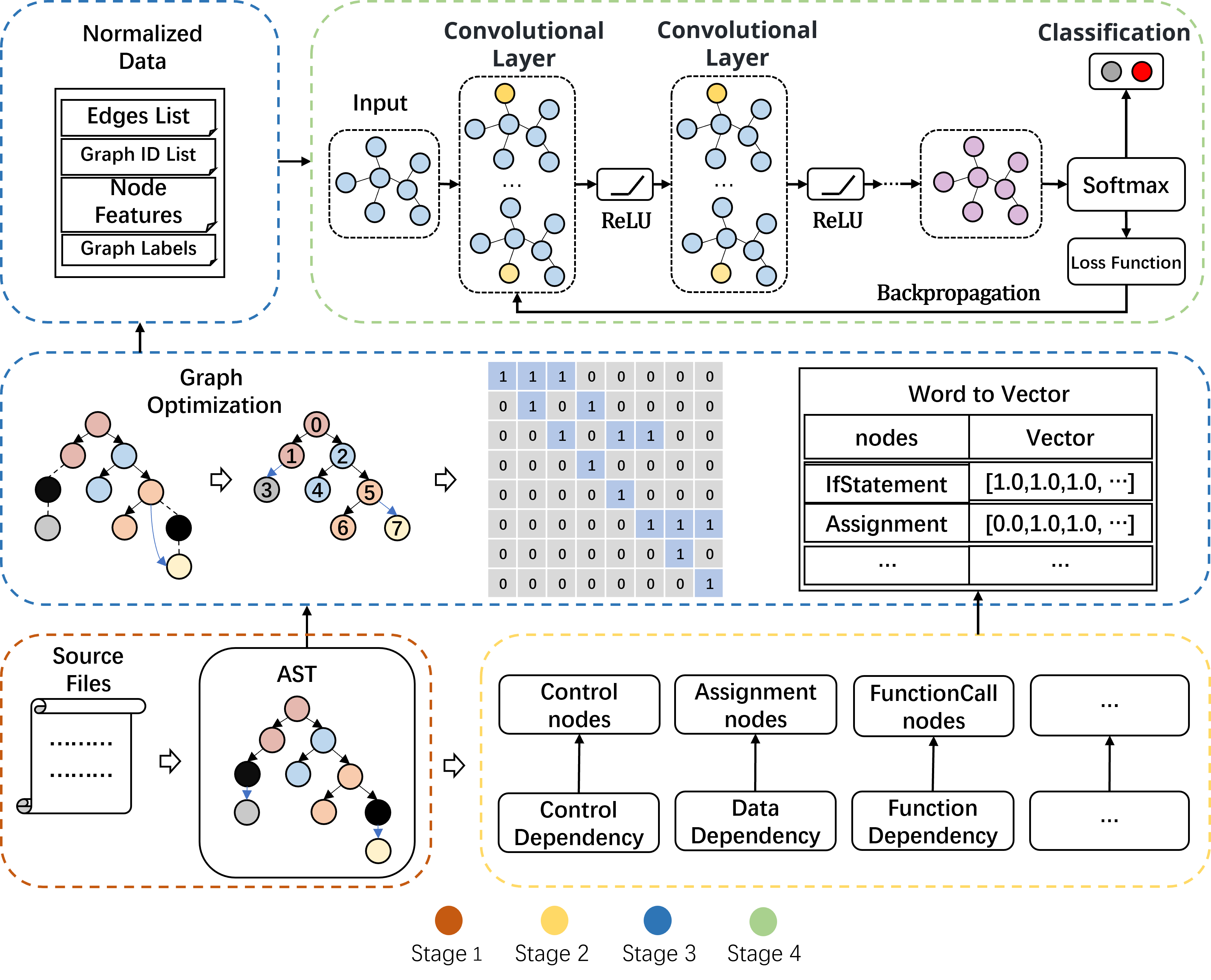}
\vspace{-2.5em}
\caption{Overview of the Framework.}
\label{fig:module}
\vspace{1em}
\end{figure}

According to Fig. \ref{fig:module}, the overall architecture of our approach consists of four stages:
\underline{(1)} AST Generation stage: This stage compiles the source code to generate a complete AST.
\underline{(2)} Feature Extraction stage: The critical features of the smart contract are extracted by parsing the AST. 
\underline{(3)} Graph Processing stage:  This stage represents and optimizes the extracted data in a graphical structure. Then, the graph and features are transformed into normalized data that GCN can process.
\underline{(4)} Defect Detection Based on GCN: This stage uses GCN to process normalized data, identify patterns, and learn features for smart contract analysis.
Algorithm \ref{alg:alg1} outlines the entire processing flow.

\textbf{AST Generation stage:} We can thoroughly analyze the code to discover and optimize security defects by converting smart contract code into an AST~\cite{Wang_2022_Unified}. 
Smart contract code can be thoroughly analyzed for security defects and optimized by converting it to an AST.
The AST employs depth and branching to represent the complexity and decision-making constructs within the code. 
By traversing the AST, We can extract crucial features and allocate different roles to different program elements to build a contract graph~\cite{Zhuang_2020_Smart}.

\setlength{\textfloatsep}{0pt}
\begin{algorithm}[!t]
\caption{Source Code to Normalized Data}
\footnotesize
\label{alg:alg1}
\begin{algorithmic}[1]
\Procedure{SourceToGraph}{$source\_file$}
    \State $AST \gets \text{Parse}(source\_file)$
    \State $word2idx, M \gets \text{Preprocess}(AST)$
    \State $A, N, V \gets \text{ASTtoAdjMatrixAndDict}(AST, word2idx)$
    \State $G \gets \text{OptimizeGraph}(A, N, V)$
    \State $G' \gets \text{Normalize}(G)$
    \State \Return $G'$
\EndProcedure

\Procedure{ASTtoAdjMatrixAndDict}{$AST, word2idx$}
    \State $A, N, V \gets \text{Initialize empty matrix and dictionaries}$
    \For{each $node$ in $AST$}
        \State $w_i \gets \text{node.attributes}$
        \State $x_i \gets M[:, word2idx(w_i)]$
        \State $V[i] \gets x_i$
        \For{each $neighbor$ in $node.neighbors$}
            \State $A[i, word2idx(neighbor.label)] \gets 1$
            \State $N[i] \gets neighbor$
        \EndFor
    \EndFor
    \State \Return $A, N, V$
\EndProcedure

\Procedure{OptimizeGraph}{$A, N, V$}
    \State $V' \gets \text{RemoveNodesAndEdges}(A, N, V)$
    \State $G \gets \text{DFSAndRemove}(V')$
    \State \Return $G$
\EndProcedure

\Procedure{Normalize}{$G$}
    \State $G' \gets \text{ApplyTransformations}(G)$
    \State \Return $G'$
\EndProcedure
\end{algorithmic}
\end{algorithm}

\textbf{Feature Extraction stage:}  Syntactic features of the defective code can extract matching code fragments through data structures~\cite{AlDebeyan_2022_Improving}. 
In defect detection tasks, node features are crucial to represent the structure of the code graph, and their importance depends on the node type. 
\textbf{Declaration dependency nodes} can represent the code's input, output, and state variables.
\textbf{Expression dependency nodes} encapsulate program logic and computation.
\textbf{Control dependency nodes}  define the execution flow of a program.
\textbf{Data dependency nodes} refers to the dependence of certain program parts on the state or output of other parts. 
\textbf{function dependency nodes} focus on the relationships between functions, which is critical to understanding how other functions affect the behavior of one function. 
We use a set $L$ to capture critical features for defect detection by focusing on crucial nodes and their attributes represented by tuples $(N_{id}, N_n, N_t, N_v)$ or $w$, including the node's unique ID, name, type, and potential value, for efficient and immutable data representation.  Directed edges are represented by tuples $(E_s,E_e,E_t)$ labeled with the start node, end node, and edge type, thus simplifying the representation of potential paths and enabling efficient analysis.

\textbf{Graph Processing stage:} We enhance graph-based smart contract analysis by optimizing the construction and processing of graphs~\cite{Zhuang_2020_Smart}. 
First, we use depth-first traversal to construct the graph starting from the AST root node, creating nodes and edges that reflect the structure of the contract.
 Then, the graph is optimized by pruning nodes unrelated to the predefined label set $L$, streamlining the structure for more efficient analysis.
 Next, the nodes are converted into feature vectors using the $word2idx$ dictionary~\cite{Asudani_2023_Impact} and embedding matrix. 
This embedding matrix, adjacency matrix, and dictionary encapsulate the graph's topology. 
Finally, we normalize the graph by refining the nodes into a set of feature vectors and capturing the connections using the adjacency matrix.

\textbf{Defect Detection Based on GCN:}
In GCN, node features update by aggregating neighboring features as follows:
\vspace{-0.5em}
$$ H^{(l+1)} = \sigma\left(\hat{D}^{-1/2} \hat{A} \hat{D}^{-1/2} H^{(l)} W^{(l)}\right) $$
\vspace{-1.5em}

where $ \hat{A} = A + I_N $ is the adjacency matrix with added self-loops, $ \hat{D} $ is the degree matrix, $ H^{(l)} $ is the feature matrix at layer $ l $, $ W^{(l)} $ is the weight matrix, and $ \sigma $ is the activation function. This process helps feature learning by aggregating and transforming node features through multiple layers.

\vspace{-0.8em}
\section{Experiment} \label{sec:experiment}
\vspace{-1ex}

All experiments are executed on a server equipped with NVIDIA GeForce GTX 4070Ti GPU, Intel(R) Core(TM) i9-13900KF CPU, and 128G RAM, operating on Ubuntu 22.04 LTS. The software environment includes Python 3.9 and PyTorch 2.0.1.

\textbf{Dataset.}
In this study, we utilize the DAppSCAN dataset~\cite{zheng2023dappscan} to provide insights into the defects of DEX smart contracts. The dataset includes 703 different projects and 23,637 smart contracts. 
We specifically focus on 46 DEX projects that contain 5,671 smart contracts. 
In addition, we analyzed 1,311 security audit reports provided by 30 different entities and performed a cross-reference analysis of these reports with the relevant smart contracts. 
The experiments also use the Smartbugs dataset~\cite{DurieuxEtAl2020ICSE}, which contains 4,285 smart contract codes with known defects. 
The datasets form the basis for our experimental analysis. Table \ref{table:dataset} summarizes the smart contract data used.

\begin{table}[h!]
\centering
\small
\vspace{-1em}
\caption{The Collected Dataset for Our Evaluation. \# indicates the number of each item.}
\vspace{-1.5em}
\begin{tabular}{>{\centering\arraybackslash}p{2.2cm} c c}
\toprule
\textbf{Dataset} & \textbf{\# Contracts} & \textbf{ \# Audit Reports} \\
\midrule
DAppSCAN & 5,671 & 1,311 \\
\hline
Smartbugs & 2,000 & 0 \\
\bottomrule
\end{tabular}
\label{table:dataset}
\vspace{-1.5em}
\end{table}

\textbf{Evaluation Metrics.} The \textsc{StateGuard} is evaluated based on the following research questions (RQs):
\underline{RQ1:} Is \textsc{StateGuard} capable of accurately identifying state derailment defects in the public dataset?
\underline{RQ2:} Can \textsc{StateGuard} find state-related defects undetectable by other tools? How does it compare with existing tools?
\underline{RQ3:} Can \textsc{StateGuard} effectively detect defects in real-world contracts?

\textbf{Answer to RQ1: Defects Detection in a Large-Scale Dataset.} 
In the DAppSCAN dataset, we experimentally analyzed 5,671 smart contracts. We adopted a common data partitioning strategy where 90\% of the data is used for model training while the remaining 10\% is set aside for testing. The experimental results are presented in Table \ref{tab:table1}, demonstrating \textsc{StateGuard}'s smart contract defect identification performance, including Accuracy (Acc), Recall, Precision, F1 Score, and False Positive Rate (FPR). 
It is important to note that the goal of \textsc{StateGuard} is to determine whether a smart contract contains a defect, not its specific number of occurrences. Therefore, even if the same defect occurs multiple times in a contract, we only count it as once.
\begin{table}[htbp] 
\vspace{-1.2em}
\centering 
\small
\caption{Performance Metrics of StateGuard.}
\vspace{-1em}
\label{tab:table1} 
\begin{threeparttable}
\begin{tabular}{@{}lccccc@{}}
\toprule 
\textbf{Tool} & \textbf{Acc(\%)} & \textbf{Recall(\%)} & \textbf{Precision(\%)} & \textbf{F1(\%)} & \textbf{FPR(\%)} \\
\midrule 
StateGuard & 94.83 & 94.82 & 98.28 & 94.25 & 0.03 \\
\bottomrule 
\end{tabular} 
\end{threeparttable}
\vspace{-1.5em}
\end{table}

The results show that \textsc{StateGuard}'s accuracy is as high as 94.83\%, and its recall rate is also 94.82\%, which shows that it can accurately identify state derailment defects. The precision rate of 98.28\% means that a very high percentage of the defects identified by \textsc{StateGuard} are real. The F1 score of 94.25\% comprehensively reflects its good performance. The FPR is only 0.03\%, further demonstrating \textsc{StateGuard}'s low error rate in misclassifying non-defective contracts as defective.

In summary, \textsc{StateGuard}'s performance results in identifying state derailment defects on public datasets, demonstrating its reliability and effectiveness in DEX contract security analysis.

\textbf{Answer to RQ2: Comparison Experiment.} 
In existing research, we observe that most analytical tools are limited to analyzing a single smart contract, and they fail to adequately address the complexity presented by DApps involving multiple contracts.
To cross the adaptability limitations of these tools, we use the traditional smart contract dataset SmartBugs for benchmarking. 
Following Yang et al. ~\cite{Yang_2023_Definition}, 2,000 contracts were selected for the experiment. Half of these contracts are positive examples containing state derailments, and the other half are negative examples without such defects. 
For comparative analysis, we chose five tools—Mythril~\cite{mythril}, Oyente~\cite{Luu_2016_Making}, Securify~\cite{tsankov2018securify}, Confuzzius~\cite{Torres_2021_ConFuzzius} and Conkas~\cite{Veloso__Conkas}—based on their source code availability, state defect detection capabilities, and precise defect location reporting.

\begin{table}[!t]
\centering
\vspace{-1.2em}
\small
\caption{Performance Comparison of Related Tools.} 
\vspace{-1.5em}
\label{tab:performance_comparison} 
\begin{tabular}{@{}lccccc@{}} 
\toprule
\textbf{Tools} & \textbf{Acc(\%)} & \textbf{Recall(\%)} & \textbf{Precision(\%)} & \textbf{F1(\%)} & \textbf{FPR(\%)} \\ 
\midrule
Mythril     & 34.89   & 47.34      & 50.26         & 48.75        & 88.67   \\
Confuzzius  & 53.43   & 53.44      & 66.03         & 59.07        & 46.59   \\
Oyente      & 52.53   & 50.15      & 90.67         & 64.58        & 32.48   \\
Securify    & 74.10    & 56.90       & 86.74         & 68.72        & 8.70     \\
Conkas      & 74.72   & 81.10       & 89.34         & 85.02        & 74.13   \\
\textbf{StateGuard} & \textbf{91.40}    & \textbf{90.40}       & \textbf{92.24}         & \textbf{91.31}       & \textbf{7.60}     \\
\bottomrule
\end{tabular}
\end{table}

To fairly evaluate the performance of each tool, we filtered the analyzed results and kept only the valid results.
Table \ref{tab:performance_comparison} shows that \textsc{StateGuard} outperforms other detection tools in all five key performance metrics. Specifically, \textsc{StateGuard} achieved 91.40\% accuracy, 90.40\% recall, 92.24\% precision, and 91.31\% F1 score while maintaining a low FPR of 7.60\%. 

In summary, \textsc{StateGuard} outperforms other tools in detecting state-related defects in smart contracts, with higher accuracy, recall, precision, and F1 score while maintaining a lower FPR.

\textbf{Answer to RQ3: Real-world Contract Detection.} 
We randomly select 1,596 samples of smart contracts from Etherscan~\cite{etherscan}, which cover smart contracts of different sizes. The sampling methodology we adopted ensures the broad applicability and validity of the findings.

We use \textsc{StateGuard} to detect these smart contracts from the real world, and the results show that \textsc{StateGuard} can successfully identify contracts with state derailment defects. \textsc{StateGuard} has discovered multiple novel defects with identifiers CVE-2023-47033, CVE-2023-47034, and CVE-2023-47035. These defects have been publicized and notified to the vendor. We have also submitted a detailed security audit report to Etherscan that includes the smart contract address with the defect, the exact location of the defect, its nature, and the potential impact. Notably, the benchmark tools referenced in RQ2 failed to discover these defects. 

In summary, \textsc{StateGuard}'s success in identifying defects in real-world smart contracts proves its utility and effectiveness in handling smart contracts of varying sizes.

\vspace{-0.6em}
\section{Discussion and Conclusion}
\label{sec:conclusion}
The current graphical representation for smart contract analysis has a time complexity of $O(V + E)$, which presents a challenge when dealing with the expansive and interconnected nature of DApps' smart contracts. Although our graph optimization methods have provided some benefit, the complexity inherent in multiple AST traversals for comprehensive analyses, such as those involving data or control dependencies, persists as a challenge. Future work will aim to refine these graphical representations and involve domain experts more closely. We plan to explore Large Language Models (LLMs) for smart contract defect detection\cite{li2023overview,mao2024automated}.

We introduced \textsc{StateGuard}, a deep-learning framework for identifying state derailment defects in DEX smart contracts. textsc{StateGuard} converts the source code to an AST, optimizes it for graphs, and then uses GCN for detection. The method proves effective, with precision and recall of over 90\% on both datasets. It has also successfully identified real-world contract defects.

\vspace{-0.4em}
\section{ACKNOWLEDGMENTS}
This work is sponsored by the National Natural Science Foundation of China (No.62362021), the CCF-Tencent Rhino-Bird Open Research Fund (No.RAGR20230115), and the Hainan Provincial Department of Education Project (No.HNJG2023-10).


\vspace{-0.4em}
\normalem
\bibliographystyle{ACM-Reference-Format}
\bibliography{ref}

\end{document}